\documentclass[aps,prd,twocolumn,superscriptaddress,preprintnumbers,floatfix,nofootinbib,notitlepage]{revtex4-1}

\usepackage{graphicx}
\usepackage{amsmath,amsfonts,amssymb}
\usepackage{hyperref}
\usepackage{color}

\def\lsim{\mathrel{\raise.3ex\hbox{$<$\kern-.75em\lower1ex\hbox{$\sim$}}}}
\def\gsim{\mathrel{\raise.3ex\hbox{$>$\kern-.75em\lower1ex\hbox{$\sim$}}}}

\newcommand{\be}{\begin{equation}}
\newcommand{\ee}{\end{equation}}
\newcommand{\bea}{\begin{equation}\begin{aligned}}
\newcommand{\eea}{\end{aligned}\end{equation}}
\newcommand{\td}{{\rm d}}

\newcommand*{\Scale}[2][4]{\scalebox{#1}{$#2$}}

\begin{document}

\preprint{KCL-PH-TH/2020-15, CERN-TH-2020-048}

\title{Probes of Gravitational Waves with Atom Interferometers}

\author{John~Ellis}
\email{john.ellis@cern.ch}
\affiliation{Physics Department, King's College London, London WC2R 2LS, UK}
\affiliation{Theoretical Physics Department, CERN, Geneva, Switzerland}
\affiliation{National Institute of Chemical Physics \& Biophysics, R\"avala 10, 10143 Tallinn, Estonia}

\author{Ville Vaskonen}
\email{ville.vaskonen@kcl.ac.uk}
\affiliation{Physics Department, King's College London, London WC2R 2LS, UK}
\affiliation{National Institute of Chemical Physics \& Biophysics, R\"avala 10, 10143 Tallinn, Estonia}

\begin{abstract}
Atom interferometers (AIs) on earth and in space offer good capabilities for measuring gravitational waves (GWs) in the mid-frequency deciHz band, complementing the sensitivities of the LIGO/Virgo and LISA experiments and enabling probes of possible modifications of the general relativity predictions for GW propagation. We illustrate these capabilities using the projected sensitivities of the AION (terrestrial) and AEDGE (space-based) AI projects, showing that AION could improve the present LIGO/Virgo direct limit on the graviton mass by a factor $\sim 40$ to $\simeq 10^{-24}$~eV, and AEDGE could improve the limit by another order of magnitude. AION and AEDGE will also have greater sensitivity than LIGO to some scenarios for Lorentz violation.
\end{abstract}

\maketitle

\section{Introduction}

The discovery of gravitational waves (GWs) by the LISA and Virgo laser interferometer (LI) experiments~\cite{LIGOVirgo} has opened new perspectives in astrophysics, cosmology and fundamental physics. In particular, new tests of general relativity (GR) and its predictions for the emission and propagation of GWs have become possible. The LIGO/Virgo breakthrough will be followed by several other approved LI experiments (KAGRA~\cite{KAGRA}, INDIGO~\cite{INDIGO} and LISA~\cite{LISA}), with other terrestrial and space LI experiments being proposed. The ground-based experiments typically have maximal sensitivities at frequencies $f \gtrsim 10$~Hz, while space-based experiments typically have maximal sensitivities for $f \lesssim 0.1$~Hz.

Atom interferometers (AIs) are candidates for making GW measurements in the mid-frequency deciHz gap between LIGO/Virgo and LISA, where they would complement the LI experiments, e.g., by extending the durations of observations of mergers by these and other LI experiments, and possibly observing the mergers of intermediate-mass black holes (BHs). MAGIS~\cite{MAGIS} is a pioneering terrestrial AI experiment being prepared in the US, AION~\cite{AION} is a similar project in the UK, proposing a series of detectors with lengths $\sim 10$\,m, $\sim 100$\,m and $\sim 1$\,km, respectively, and AEDGE~\cite{AEDGE} is a concept for a space-based AI experiment in the longer term.~\footnote{Other AI projects for exploring gravitational physics have also been proposed~\cite{MIGA,ELGAR,ZAIGA}.}

The capabilities of AION and AEDGE for observing BH mergers and possible cosmological sources of GWs such as first-order phase transitions in the early universe and cosmic strings have been documented in~\cite{AION} and~\cite{AEDGE}, respectively. Their capabilities for tests of fundamental physics were are also mentioned, but not explored in detail. On the other hand, there have been evaluations of the capabilities of other proposals for experiments in the mid-frequency deciHz range such as DECIGO~\cite{DECIGO}, including the possibilities of joint analyses with other experiments such as LISA and the proposed Einstein Telescope (ET~\cite{ET} and Cosmic Explorer (CE)~\cite{Carson:2019kkh}).

In this paper we complement the previous studies of AION~\cite{AION} and AEDGE~\cite{AEDGE} by exploring their stand-alone capabilities for constraining possible deviations from GR in the propagation of GWs from BH mergers, including events similar to those measured by LIGO/Virgo and possible measurements of mergers involving heavier BHs. We focus on the possible constraints on the graviton mass, $m_g$, that could be established by AION and AEDGE, and also discuss their possible constraints on Lorentz-violating (LV) modifications of GW propagation~\cite{Will:1997bb,Mirshekari:2011yq,Carson:2019kkh}.~\footnote{We concentrate on the capabilities of AEDGE and the 1\,km stage of AION, with some remarks on AION 100\,m.} 

Prospective AION and AEDGE measurements offer two advantages over the LIGO/Virgo measurements that have already set a 90\% CL direct upper limit $m_g < 4.7 \times 10^{-23}$~eV~\cite{LIGOScientific:2019fpa}.~\footnote{For a review of other bounds on the graviton mass, see Ref.~\cite{deRham:2016nuf}.} The AI experiments offer much longer observations of the inspiral stages of events of the types measured by LIGO/Virgo, and possibly mergers of heavier BHs that emit GWs of lower frequencies, which have enhanced sensitivity to $m_g$. We find that measurements of an event similar to the LIGO/Virgo discovery event GW150914 with the 1\,km stage of AION could improve the 90\% CL direct upper limit to $m_g < 1.1 \times 10^{-24}$~eV, and AEDGE measurements could further improve it to $m_g < 1.3 \times 10^{-25}$~eV. Further improvements in the sensitivity to $m_g$ could come from measurements of the mergers of more massive BHs. We find also that AION 1\,km and AEDGE will have greater sensitivity than LIGO to LV by amounts $\propto A^\alpha$ with $\alpha \leq 1$.

\section{Analysis}

We consider the potential GW signal from a BH-BH binary with component masses $m_1$ and $m_2$. The Fourier transform of the waveform from the binary inspiral is given by
\be
\tilde{h}(f) = A(f) \,e^{i\Psi(f)} \,,
\ee
where $A(f)$ is the amplitude and $\Psi(f)$ is the phase of the strain $h$ due to the GWs emitted during the quasi-circular inspiral. To calculate the waveform we use the {\tt PhenomD} model~\cite{Khan:2015jqa}, in which spin-independent corrections are included up to 3.5\,PN order~\cite{Buonanno:2009zt,Blanchet:2013haa}, linear spin-orbit corrections up to 3.5\,PN order~\cite{Bohe:2013cla}, and quadratic spin corrections up to 2\,PN order~\cite{Poisson:1997ha,Arun:2008kb,Mikoczi:2005dn}. We take into account the time dependence of the observed signal arising from the motion of the detector around the Earth and the Sun, following Ref.~\cite{Graham:2017lmg}. The free parameters are the source direction $\hat{n} = \hat{n}(\theta,\phi)$ and luminosity distance $D_L$, the binary mass ratio $q \equiv m_1/m_2$ and the chirp mass $\mathcal{M}_z \equiv (1+z)(m_1 m_2)^{3/5}/(m_1+m_2)^{1/5}$, the coalescence time $t_c$, the binary orbit inclination $\cos\tau_i$, the polarization angle $\psi$ and phase $\phi_c$ of the signal at $t_c$, and the symmetric and anti-symmetric dimensionless spin parameters, $\chi_{s,a} \equiv (\chi_1\pm \chi_2)/2$.

\begin{figure}
\centering
\includegraphics[width=0.49\textwidth]{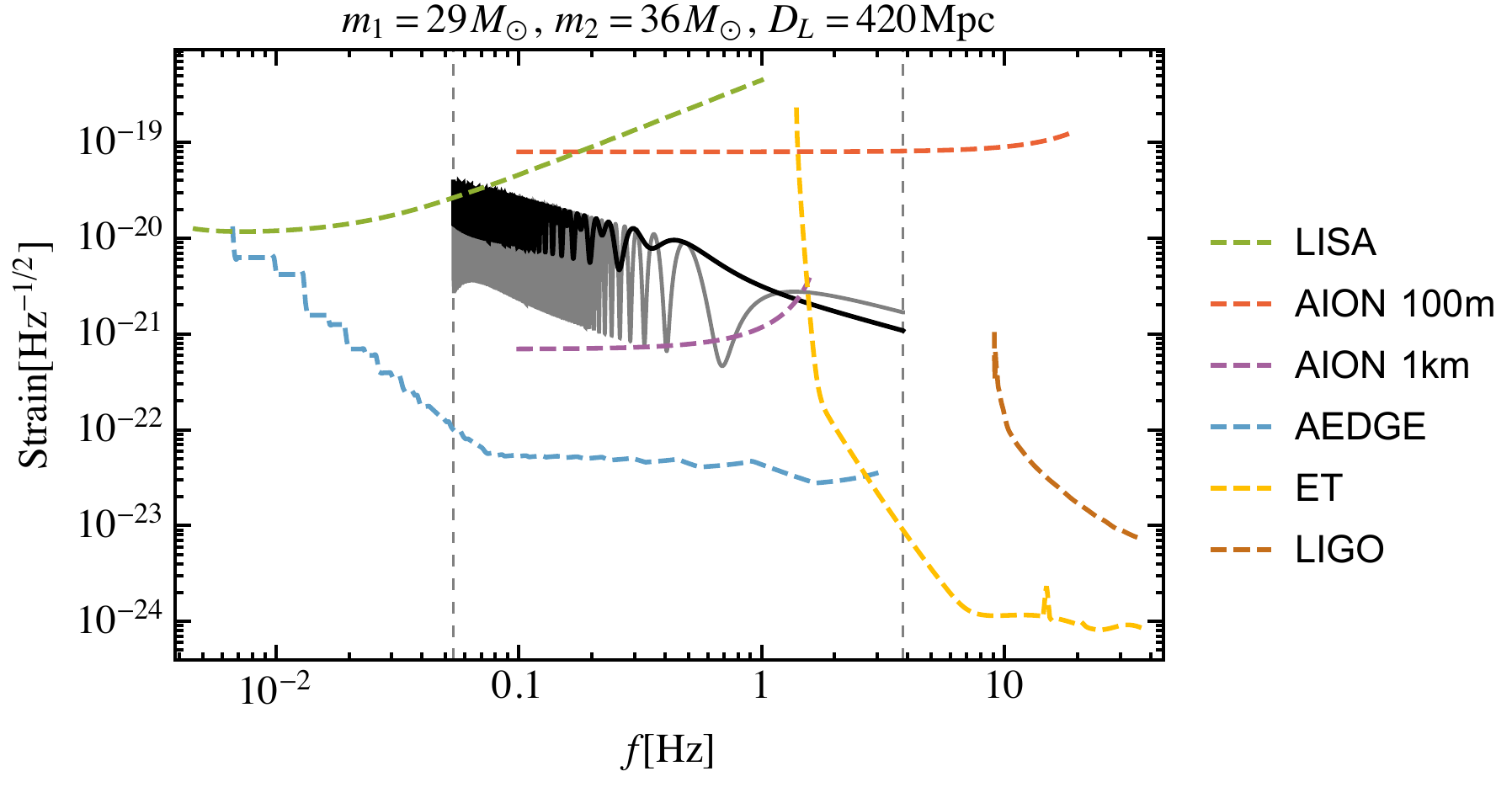}
\caption{\it The solid lines show the GW strain from a GW150914-like BH-BH binary inspiral calculated for terrestrial (black) and space-based (grey) detectors. The dashed curves show the expected detector noises of LIGO and various future GW detectors. The assumed source component masses and luminosity distance are given at the top of the plot. Its direction and the AION and AEDGE detector configurations are described in the text. The left and right vertical dashed lines indicate the frequency 60\, days and 1\,minute before the binary merges, where we cut the signal.}
\label{fig:strain}
\end{figure}

An example of the GW strain $2\sqrt{f}|\tilde h(f)|$ from a BH-BH binary inspiral is shown together with the noise curves of LIGO and future GW detectors in Fig.~\ref{fig:strain}. The signals are calculated separately for ground- and space-based detectors, and exhibit oscillations due to the changing {orientations} of the detectors. In this example we use $\theta = 100^{\circ}$, $\phi = 30^{\circ}$, $\psi = 60^{\circ}$, $\phi_c = 0$ and $\tau_i = 150^{\circ}$, and binary parameters similar to the GW150914 event~\cite{TheLIGOScientific:2016wfe}, with component spins $\chi_1 = \chi_2 = 0$. We assume that the ground-based detector is oriented along an Earth radial direction and is located at a latitude of $46.2^{\circ}$ (similar to that of CERN). The space-based detector consists of two satellites in identical circular geocentric orbits with an inclination of $28.5^{\circ}$ at an altitude of $24\times10^3$\,km from the Earth's centre, forming a baseline of $44\times10^3$\,km and completing each orbit in 10\,hours~\cite{Graham:2017pmn}.

\begin{figure}
\centering
\includegraphics[width=0.49\textwidth]{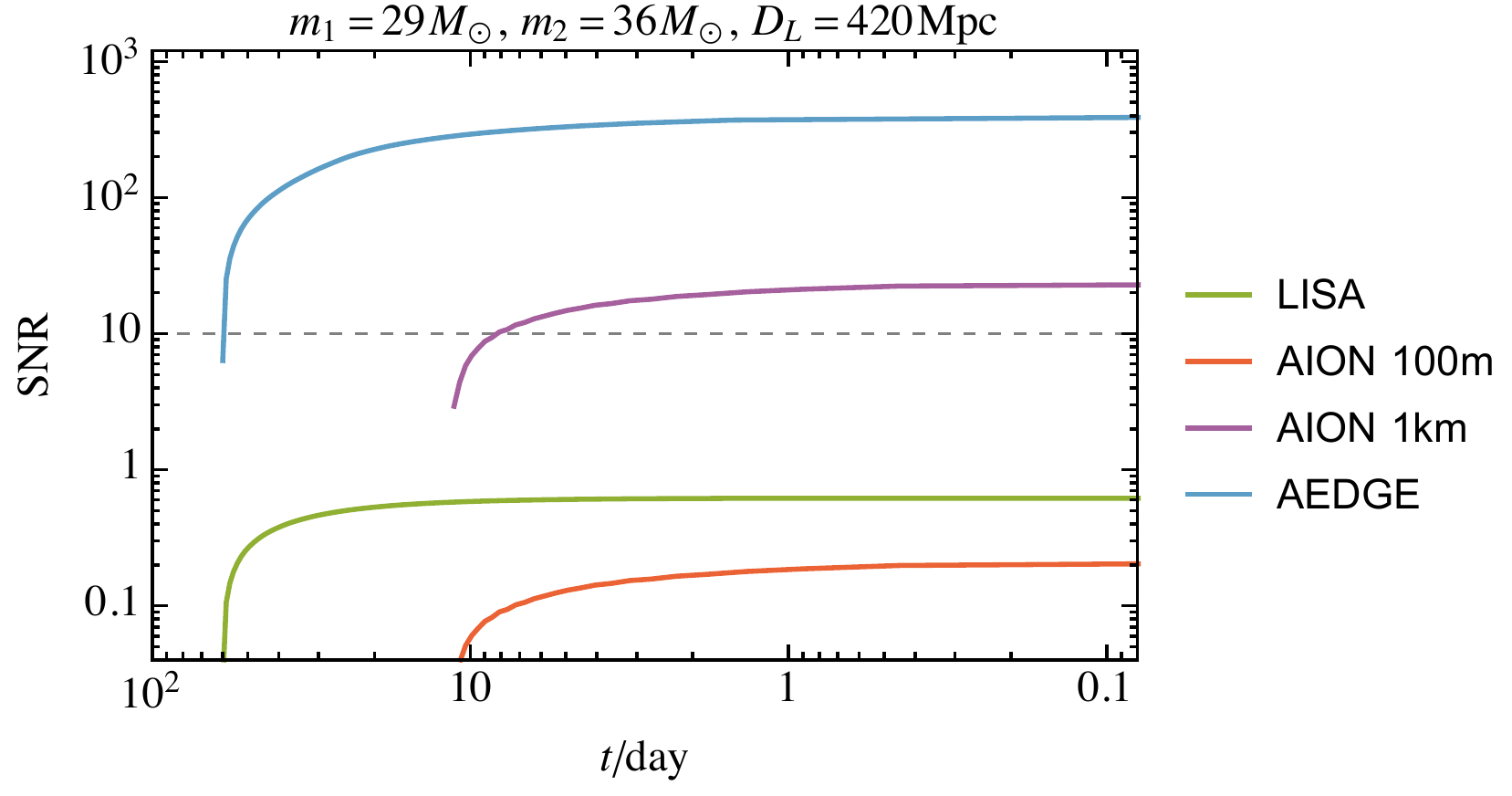}
\caption{\it The signal-to-noise ratio (SNR) for the signal shown in Fig.~\ref{fig:strain} as seen in different detectors as a function of time during the last 60\,days before the merger.}
\label{fig:SNR1}
\end{figure}

\begin{figure}
\centering
\includegraphics[width=0.49\textwidth]{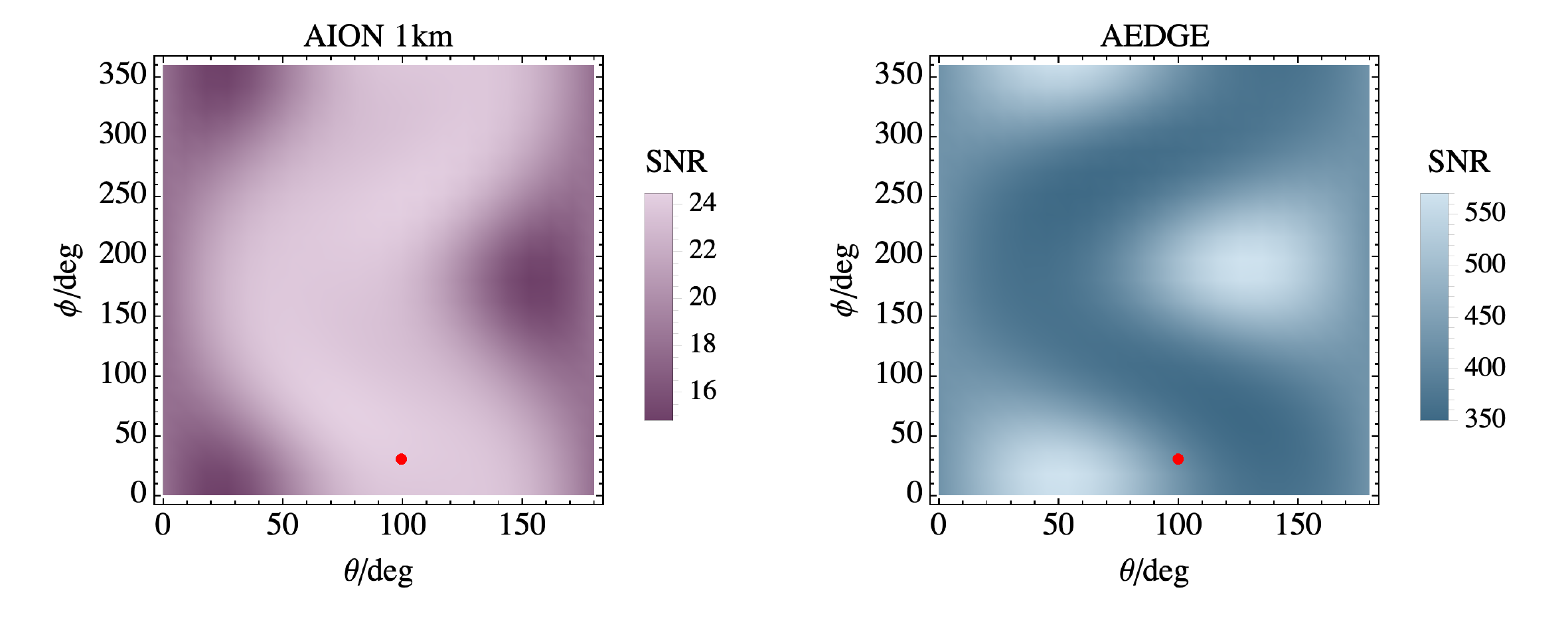}
\includegraphics[width=0.49\textwidth]{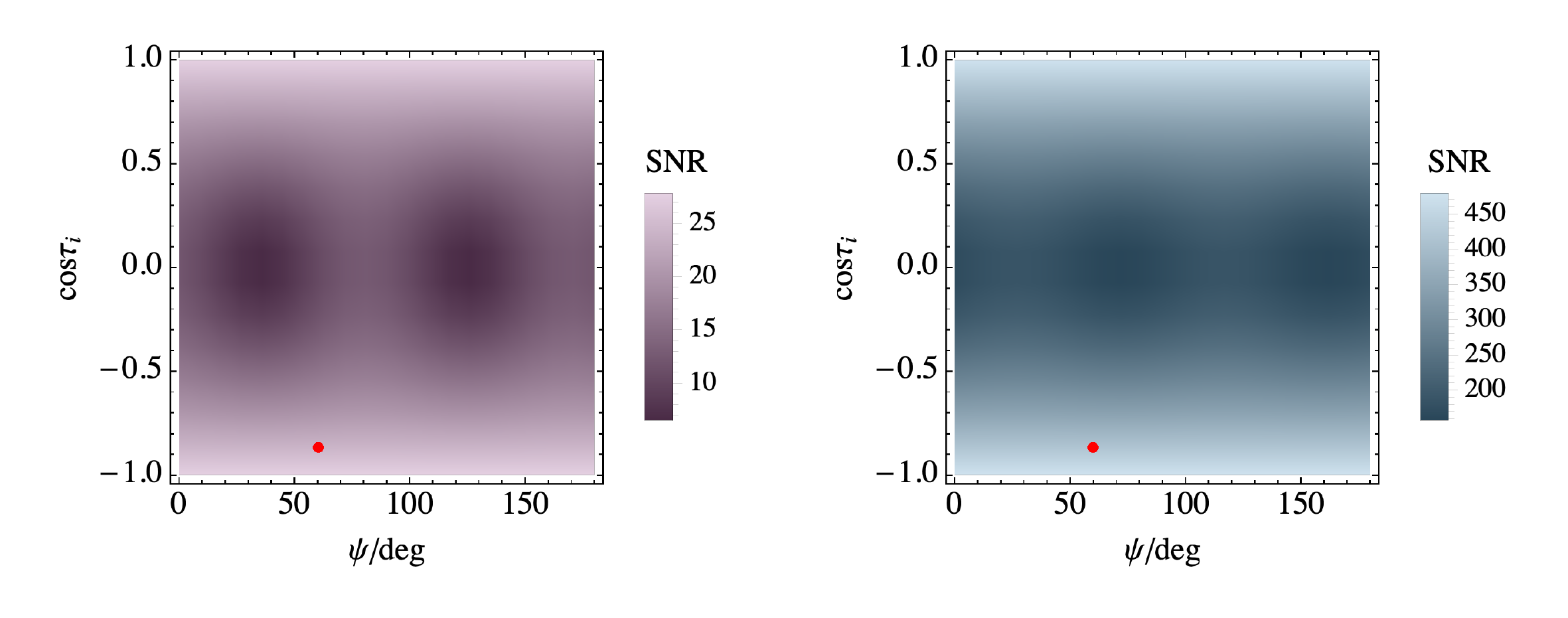}
\caption{\it The dependences of the SNR in AION 1\,km (left panels) and AEDGE (right panels) on the direction $(\theta, \phi)$ of the source (upper panels), and on the polarization $\psi$ of the signal and the binary orbit inclination $\cos\tau_i$ (lower panels) are colour-coded, and the parameters of the benchmark GW150914-like source are indicated by the red point.}
\label{fig:SNR2}
\end{figure}

\begin{table*}
\centering
\begin{tabular}{ p{2.cm} | p{1.8cm} p{1.1cm} p{1.1cm} p{1.5cm} p{1.2cm} p{1.2cm} p{1.2cm} p{1.2cm} p{1.2cm} p{1.1cm} p{1.1cm}} \hline\hline
 & $\sigma_{\mathcal{M}_z}/M_\odot$ & $\sigma_q$ & $\sigma_{t_c}/{\rm s}$ & $\sigma_{D_L}/{\rm Mpc}$ & $\sigma_\theta/{\rm deg}$ & $\sigma_\phi/{\rm deg}$ & $\sigma_\psi/{\rm deg}$ & $\phi_c/{\rm deg}$ & $\sigma_{\cos\tau_i}$ & $\sigma_{\chi_a}$ & $\sigma_{\chi_s}$ \\ \hline
AION 1\,km & $6.7\times 10^{-4}$ & $0.42$ & $12$ & $300$ & $2.7$ & $2.7$ & $150$ & $-$ & $0.67$ & $0.99$ & $0.13$ \\
AEDGE & $1.1\times 10^{-5}$ & $0.010$ & $0.38$ & $14$ & $0.15$ & $0.077$ & $7.7$ & $77$ & $0.031$ & $0.80$ & $0.10$ \\
\hline\hline
\end{tabular}
\caption{\it {Standard deviations of the GW waveform parameters} for the AION 1\,km and AEDGE signals shown in Fig.~\ref{fig:strain}, assuming no modification of GR, for the GW150914-like benchmark source with $\mathcal{M}_z = 30.7\,M_\odot$, $q = 0.80$ and $D_L = 420\,$Mpc.}
\label{table:gr}
\end{table*}

We show in Fig.~\ref{fig:SNR1} the signal-to-noise ratio (SNR), given by
\be
{\rm SNR} = \sqrt{ 4\, \int \td f\, \frac{\tilde h^*(f) \tilde h(f)}{S_n(f)}} \,,
\ee
of the signal observed from 60\,days before the binary merges or when the signal enters the sensitivity window of the detector, if that happens later.~\footnote{For example, the signal shown in Fig.~\ref{fig:strain} enters the AION sensitivity window 11.6 days before the merger, whereas it enters the AEDGE sensitivity window 44 years previously.} Notice that we consider only the inspiral phase. If the merger happens in the sensitivity window of the detector it can significantly increase the SNR.~\footnote{In this case one could use numerical relativity fits for the waveform~\cite{Khan:2015jqa} in order to include the merger and ringdown stages in our analysis.} The signal shown in Fig.~\ref{fig:strain} exits the AION 1\,km sensitivity window $7$~minutes and the AEDGE sensitivity window $2$~minutes before the binary merges, and we
do not include in our analysis possible subsequent measurements at higher frequencies by other detectors.

The upper panels of Fig.~\ref{fig:SNR2} show the dependences of the SNR from a GW150914-like source measured in AION and AEDGE on the direction $(\theta, \phi)$ of the source, and the lower panels show the dependences on the polarization $\psi$ of the signal and the binary orbit inclination $\cos\tau_i$. These plots are colour-coded as indicated, and the parameters of the benchmark GW150914-like source are indicated by the red point. We note that these parameters are relatively favourable for both AION and AEDGE.

We perform a Fisher matrix analysis in order to estimate the accuracies with which the parameters of the waveform can be measured~\cite{Finn:1992xs,Cutler:1994ys,Poisson:1995ef}. The covariance matrix is given by the inverse of the Fisher matrix:
\be
\Gamma_{ij} = 4\, {\rm Re} \int \td f\, \frac{(\partial_i \tilde h^*) (\partial_j \tilde h)}{S_n(f)} \,,
\ee
and the standard deviations of the parameter measurements $\sigma_j$ and their correlation matrix $\rho_{ij}$ are then given by
\be
\sigma_j = \sqrt{\Gamma_{jj}^{-1}}\,, \quad
\rho_{ij} = \frac{\Gamma_{ij}^{-1}}{\sigma_i \sigma_j} \,.
\ee
We impose priors $|\chi_{a,s}| < 1$ on the spin parameters by adding $(\sigma_0)^{-2} = 1$ to the $\chi_{a,s}$ diagonal elements of the Fisher matrix~\cite{Poisson:1995ef}.

In order to provide a baseline for our analysis, we first apply the Fisher matrix analysis to the prospective AION 1\,km and AEDGE signals shown in Fig.~\ref{fig:strain} in the absence of any modification of GR. We find the following correlation matrix for AION 1\,km in the $\{\mathcal{M}_z, q, t_c, D_L, \theta, \phi, \psi, \phi_c, \cos\tau_i, \chi_a, \chi_s\}$ basis:
\be
\rho = \Scale[0.64]{\left(
\begin{array}{ccccccccccc}
 1. & 0.99 & 0.07 & 0.07 & 0.05 & 0.04 & 0.01 & 0.79 & 0.07 & 0.02 & 0.13 \\
 0.99 & 1. & 0.05 & 0.03 & 0.02 & 0.09 & 0. & 0.81 & 0.03 & 0.03 & 0.11 \\
 0.07 & 0.05 & 1. & 0.31 & 0.27 & 0.96 & 0.11 & 0.07 & 0.32 & 0.01 & 0. \\
 0.07 & 0.03 & 0.31 & 1. & 0.02 & 0.33 & 0.04 & 0.02 & 1. & 0. & 0. \\
 0.05 & 0.02 & 0.27 & 0.02 & 1. & 0.02 & 0.35 & 0.15 & 0. & 0. & 0. \\
 0.04 & 0.09 & 0.96 & 0.33 & 0.02 & 1. & 0.01 & 0.02 & 0.33 & 0. & 0.02 \\
 0.01 & 0. & 0.11 & 0.04 & 0.35 & 0.01 & 1. & 0.53 & 0.01 & 0. & 0. \\
 0.79 & 0.81 & 0.07 & 0.02 & 0.15 & 0.02 & 0.53 & 1. & 0.04 & 0.15 & 0.26 \\
 0.07 & 0.03 & 0.32 & 1. & 0. & 0.33 & 0.01 & 0.04 & 1. & 0. & 0. \\
 0.02 & 0.03 & 0.01 & 0. & 0. & 0. & 0. & 0.15 & 0. & 1. & 0.99 \\
 0.13 & 0.11 & 0. & 0. & 0. & 0.02 & 0. & 0.26 & 0. & 0.99 & 1. \\
\end{array}
\right)} \,,
\ee
and the following for AEDGE:
\be
\rho = \Scale[0.64]{\left(
\begin{array}{ccccccccccc}
 1. & 0.92 & 0.83 & 0.22 & 0.32 & 0.68 & 0.03 & 0.16 & 0.21 & 0.26 & 0.26 \\
 0.92 & 1. & 0.71 & 0.15 & 0.14 & 0.55 & 0.04 & 0.51 & 0.14 & 0.6 & 0.6 \\
 0.83 & 0.71 & 1. & 0.31 & 0.5 & 0.8 & 0.11 & 0.06 & 0.29 & 0.15 & 0.14 \\
 0.22 & 0.15 & 0.31 & 1. & 0.22 & 0.24 & 0.08 & 0.09 & 1. & 0.06 & 0.06 \\
 0.32 & 0.14 & 0.5 & 0.22 & 1. & 0.04 & 0.42 & 0.09 & 0.18 & 0.01 & 0.01 \\
 0.68 & 0.55 & 0.8 & 0.24 & 0.04 & 1. & 0.13 & 0.19 & 0.24 & 0.15 & 0.15 \\
 0.03 & 0.04 & 0.11 & 0.08 & 0.42 & 0.13 & 1. & 0.2 & 0.06 & 0.01 & 0.01 \\
 0.16 & 0.51 & 0.06 & 0.09 & 0.09 & 0.19 & 0.2 & 1. & 0.09 & 0.98 & 0.98 \\
 0.21 & 0.14 & 0.29 & 1. & 0.18 & 0.24 & 0.06 & 0.09 & 1. & 0.06 & 0.06 \\
 0.26 & 0.6 & 0.15 & 0.06 & 0.01 & 0.15 & 0.01 & 0.98 & 0.06 & 1. & 1. \\
 0.26 & 0.6 & 0.14 & 0.06 & 0.01 & 0.15 & 0.01 & 0.98 & 0.06 & 1. & 1. \\
\end{array}
\right)} \,.
\ee
The corresponding measurement errors for AION 1\,km and AEDGE are shown in Table~\ref{table:gr}. We leave blank one entry where the formal uncertainty for AION 1\,km exceeds the physical range. Combining AION 1\,km or AEDGE data with measurements of the merger by LIGO/Virgo/KAGRA/INDIGO/ET/CE would reduce significantly these and other measurement uncertainties.

\section{Modified dispersion relation}

We consider simple modifications to the GW dispersion relation of the general form
\be \label{eq:dispersion}
E^2 = p^2 + A p^\alpha\,,
\ee
where $E$ is the energy and $p$ the momentum of the GW. This causes frequency-dependent dephasing of the GW waveform~\cite{Mirshekari:2011yq}: $\Psi(f) \to \Psi(f) + \delta \Psi(f)$\,, where
\be \label{eq:deltaPsi}
\delta \Psi(f) = 
\begin{cases} 
\frac{A D_\alpha(z)}{2^{2-\alpha}(\alpha-1)} \left[\pi (1+z) f \right]^{\alpha-1} \,, \quad \alpha\neq 1\,, \\
\frac{A D_1(z)}{2} \ln\left[\pi \mathcal{M}_z f\right] \,, \quad \alpha=1\,,
\end{cases}
\ee
and the distance measure $D_\alpha(z)$ is defined as
\be
D_\alpha(z) = (1+z)^{1-\alpha} \int_0^z \td z'\, \frac{(1+z')^{\alpha-2}}{H(z')} \,.
\ee
We use the Hubble expansion rate $H$ given by the Planck 2018 best fit~\cite{Aghanim:2018eyx}. For each fixed value of $\alpha$ we perform a Fisher matrix analysis as described in the previous Section where, in addition to the 10 free {GR} waveform parameters we include the modification parameter $A$. For this purpose, we also need to relate the redshift $z$, on which $\delta \Psi$ depends, to the luminosity distance
\be
D_L(z) \equiv (1+z) \int_0^z \frac{\td z'}{H(z')} \,.
\ee
Then, upon setting $A=0$, the Fisher matrix analysis gives a $90\%$ CL upper bound on $A$:
\be
A < 1.645\,\sigma_A \,.
\ee

\subsection{Graviton mass}

\begin{figure}
\centering
\includegraphics[width=0.37\textwidth]{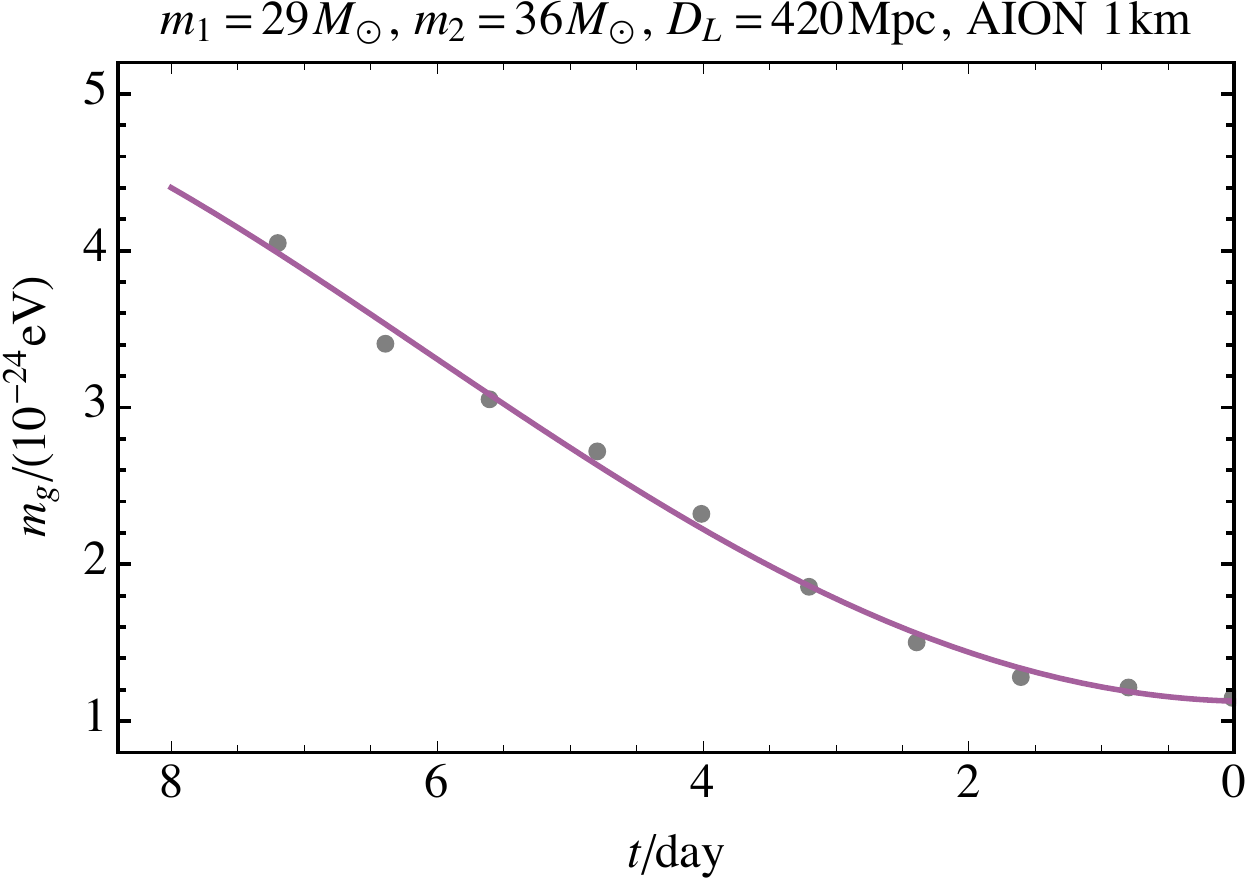}
\\ \vspace{2mm}
\includegraphics[width=0.37\textwidth]{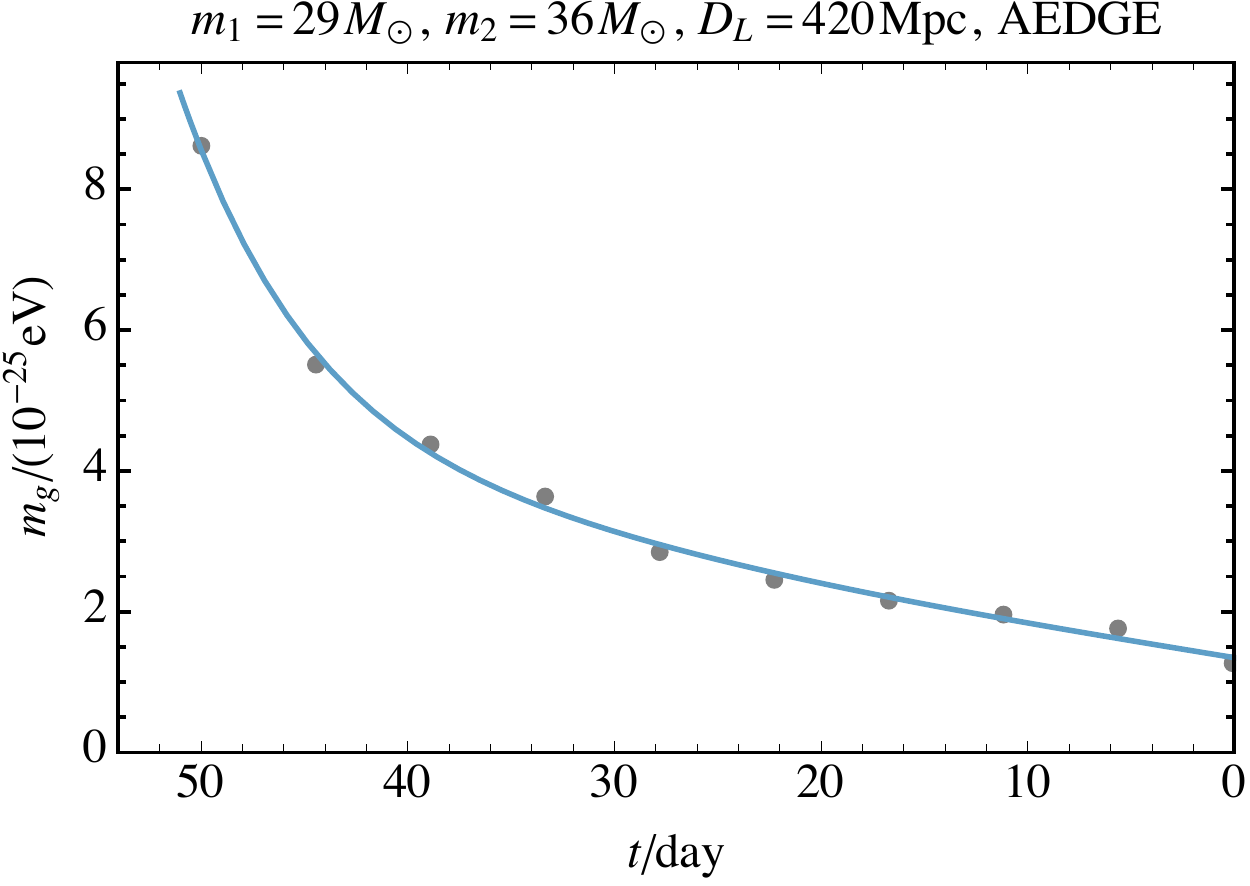}
\caption{{\it The $90\%$ upper bound on the graviton mass from GW observation of a GW150914-like BH-BH inspiral by AION 1\,km in the upper panel and by AEDGE in the lower panel. The time $t$ indicates the time until the binary merges.}}
\label{fig:mgt}
\end{figure}

The case $\alpha = 0$ is of particular interest, as it corresponds to a massive graviton, $A = m_g^2$~\cite{Will:1997bb}. The upper bound on $m_g$ obtained from observing the inspiral signal from a BH binary similar to GW150914 with AION 1\,km is shown in the upper panel of Fig.~\ref{fig:mgt} as a function of the time before the merger. {The integration of the signal is started when the signal enters the AION 1\,km sensitivity window.} Finally, when the signal exits the AION 1\,km sensitivity window, we obtain an upper bound
\be \label{AION}
m_g < 1.1\times 10^{-24}~{\rm eV} \, (90\%~{\rm CL}) \, ,
\ee
which can be compared with the current 90\% CL LIGO/Virgo upper limit $m_g < 4.7 \times 10^{-23}$~eV~\cite{LIGOScientific:2019fpa}, i.e., a factor $\sim 40$ improvement. There are large correlations in the measurement, as seen in the correlation matrix in the $\{\mathcal{M}_z, q, t_c, D_L, \theta, \phi, \psi, \phi_c, \cos\tau_i, \chi_a, \chi_s, m_g^2\}$ basis:
\be
\rho = \Scale[0.64]{\left(
\begin{array}{cccccccccccc}
 1. & 0.39 & 0.15 & 0.06 & 0.01 & 0.19 & 0. & 0.21 & 0.07 & 0.06 & 0.35 & 0.33 \\
 0.39 & 1. & 0.6 & 0.01 & 0.16 & 0.65 & 0.02 & 0.98 & 0.01 & 0.12 & 0.99 & 1. \\
 0.15 & 0.6 & 1. & 0.26 & 0.12 & 0.97 & 0.08 & 0.6 & 0.26 & 0.08 & 0.6 & 0.6 \\
 0.06 & 0.01 & 0.26 & 1. & 0.02 & 0.26 & 0.04 & 0.02 & 1. & 0. & 0.01 & 0.01 \\
 0.01 & 0.16 & 0.12 & 0.02 & 1. & 0.12 & 0.35 & 0.13 & 0. & 0.02 & 0.16 & 0.16 \\
 0.19 & 0.65 & 0.97 & 0.26 & 0.12 & 1. & 0.01 & 0.63 & 0.26 & 0.08 & 0.64 & 0.64 \\
 0. & 0.02 & 0.08 & 0.04 & 0.35 & 0.01 & 1. & 0.07 & 0.01 & 0. & 0.02 & 0.02 \\
 0.21 & 0.98 & 0.6 & 0.02 & 0.13 & 0.63 & 0.07 & 1. & 0.02 & 0.14 & 0.97 & 0.99 \\
 0.07 & 0.01 & 0.26 & 1. & 0. & 0.26 & 0.01 & 0.02 & 1. & 0. & 0.01 & 0.01 \\
 0.06 & 0.12 & 0.08 & 0. & 0.02 & 0.08 & 0. & 0.14 & 0. & 1. & 0.04 & 0.12 \\
 0.35 & 0.99 & 0.6 & 0.01 & 0.16 & 0.64 & 0.02 & 0.97 & 0.01 & 0.04 & 1. & 0.99 \\
 0.33 & 1. & 0.6 & 0.01 & 0.16 & 0.64 & 0.02 & 0.99 & 0.01 & 0.12 & 0.99 & 1. \\
\end{array}
\right)} \,.
\ee
We note in particular that the measurement of $m_g^2$ is very strongly correlated with the measurement of the mass ratio $q$, as they both contribute to the phase $\Psi$ proportionally to $(\pi \mathcal{M}_z f)^{-1}$~\cite{Will:1997bb} at leading order. Observations of the merger-ringdown signal, for example by LIGO/Virgo, KAGRA, INDIGO, ET or CE, could be used to reduce the degeneracies indicated by the correlation matrix and improve the upper bound on $m_g$. We find that if we assume that the merger time $t_c$ is known (supposing it to be detected by some other experiment), as well as the sky direction $(\theta,\phi)$ of the source as known, the upper bound on $m_g$ would improve to $m_g < 1.0\times 10^{-24}$\,eV. Adopting the procedure of Refs.~\cite{Will:1997bb,Mirshekari:2011yq,Carson:2019kkh} and averaging over the sky location $(\theta,\phi)$, binary orbit inclination $\cos\tau_i$ and polarization $\psi$, would give a slightly less stringent upper bound, $m_g < 1.3\times 10^{-24}$~eV. This difference was to be expected, since the direction of the source can be very accurately measured (see Table~\ref{table:gr}) and the sky location for our benchmark point yields a SNR that is above average (see the upper left panel of Fig.~\ref{fig:SNR2}).

\begin{figure*}
\centering
\includegraphics[width=0.8\textwidth]{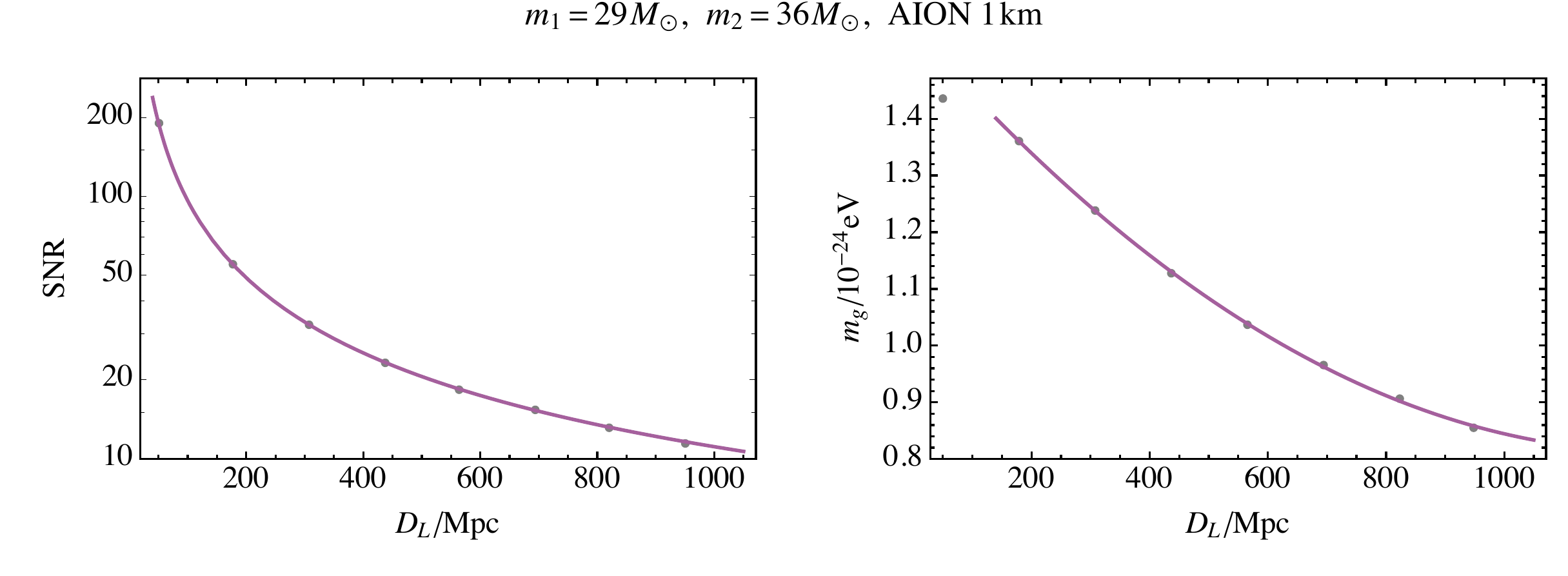}
\includegraphics[width=0.8\textwidth]{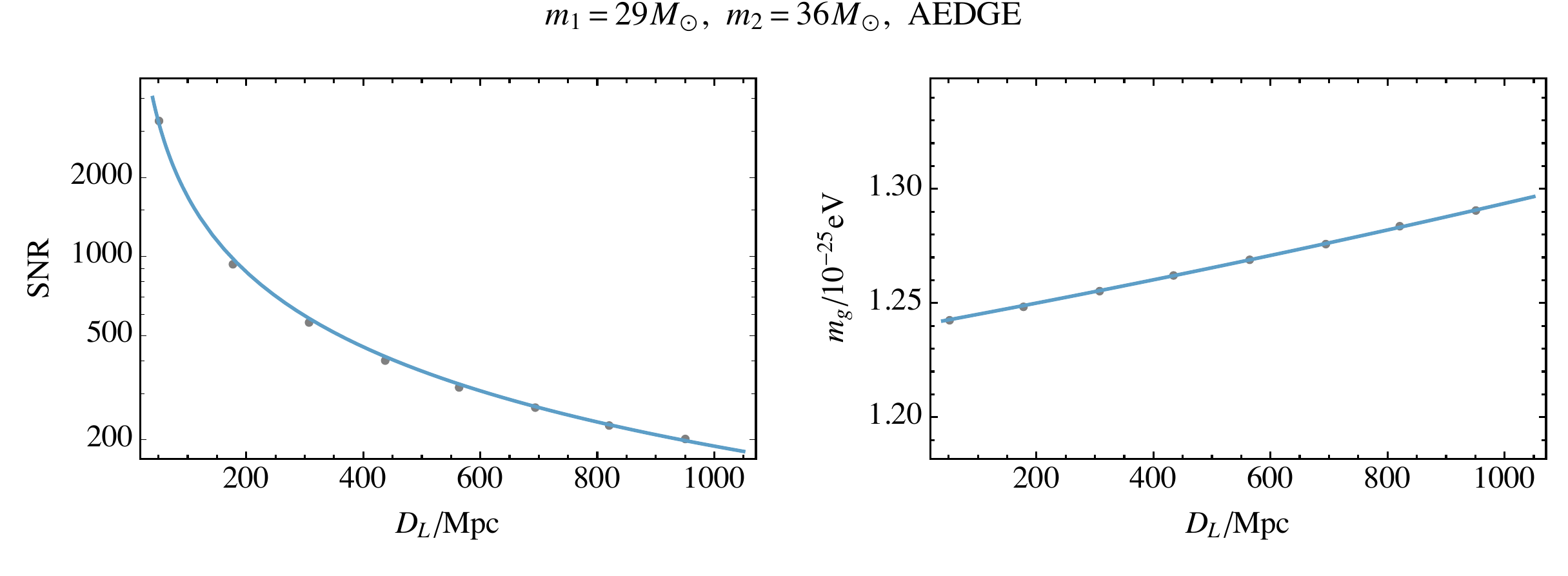}
\vspace{-5mm}
\caption{\it The signal-to-noise ratio (SNR) and the $90\%$ upper bound on the graviton mass for observations of BH-BH binary inspirals as functions of the luminosity distance for AION\,1km (upper panels) and AEDGE (lower panels).}
\label{fig:distance}
\end{figure*}

Observing the same GW150914-like signal with AEDGE for 60\,days before the binary merger would give
\be \label{AEDGE}
m_g < 1.3\times 10^{-25}~{\rm eV} \, (90\%~{\rm CL}) \, ,
\ee
i.e., another order-of-magnitude improvement.~\footnote{This sensitivity compares favourably to the combination of LISA with CE presented in~\cite{Carson:2019kkh}, and is comparable to that of the combination of DECIGO with CE as also reported there.} The upper bound as a function of the time before the merger is shown in the lower panel of Fig.~\ref{fig:mgt}. In this case the correlation matrix is
\be
\rho = \Scale[0.64]{\left(
\begin{array}{cccccccccccc}
 1. & 0.56 & 0.84 & 0.12 & 0.22 & 0.74 & 0.04 & 0.35 & 0.11 & 0.23 & 0.18 & 0.47 \\
 0.56 & 1. & 0.38 & 0.13 & 0.12 & 0.47 & 0.14 & 0.53 & 0.13 & 0.07 & 0.03 & 0.99 \\
 0.84 & 0.38 & 1. & 0.25 & 0.43 & 0.82 & 0.06 & 0.19 & 0.23 & 0.14 & 0.11 & 0.3 \\
 0.12 & 0.13 & 0.25 & 1. & 0.24 & 0.15 & 0.1 & 0.15 & 1. & 0.05 & 0.04 & 0.15 \\
 0.22 & 0.12 & 0.43 & 0.24 & 1. & 0.09 & 0.43 & 0.14 & 0.2 & 0.01 & 0.02 & 0.13 \\
 0.74 & 0.47 & 0.82 & 0.15 & 0.09 & 1. & 0.17 & 0.05 & 0.15 & 0.14 & 0.18 & 0.41 \\
 0.04 & 0.14 & 0.06 & 0.1 & 0.43 & 0.17 & 1. & 0.24 & 0.08 & 0.01 & 0. & 0.14 \\
 0.35 & 0.53 & 0.19 & 0.15 & 0.14 & 0.05 & 0.24 & 1. & 0.15 & 0.85 & 0.8 & 0.48 \\
 0.11 & 0.13 & 0.23 & 1. & 0.2 & 0.15 & 0.08 & 0.15 & 1. & 0.05 & 0.04 & 0.15 \\
 0.23 & 0.07 & 0.14 & 0.05 & 0.01 & 0.14 & 0.01 & 0.85 & 0.05 & 1. & 1. & 0. \\
 0.18 & 0.03 & 0.11 & 0.04 & 0.02 & 0.18 & 0. & 0.8 & 0.04 & 1. & 1. & 0.1 \\
 0.47 & 0.99 & 0.3 & 0.15 & 0.13 & 0.41 & 0.14 & 0.48 & 0.15 & 0. & 0.1 & 1. \\
\end{array}
\right)} \,,
\ee
showing decreases in the correlations with $m_g$ as compared to the AION 1\,km case. Assuming again that $t_c$, $\theta$ and $\phi$ are known gives a minor improvement in the upper bound on the graviton mass to $m_g < 1.2\times 10^{-25}$\,eV. Averaging over $\theta$, $\phi$, $\cos\tau_i$ and $\psi$ as in Refs.~\cite{Will:1997bb,Mirshekari:2011yq,Carson:2019kkh} would again yield a looser bound, namely $m_g < 4.0\times 10^{-25}$\,eV. In this case the difference is bigger as $\psi$ and $\cos\tau_i$ can also be very accurately measured (see Table~\ref{table:gr}), and the values for our benchmark point yield SNR that is above average (see the right panels of Fig.~\ref{fig:SNR2}). However, we expect that AEDGE would measure many similar events, enabling this average result to be improved.

\begin{figure*}
\centering
\includegraphics[width=0.8\textwidth]{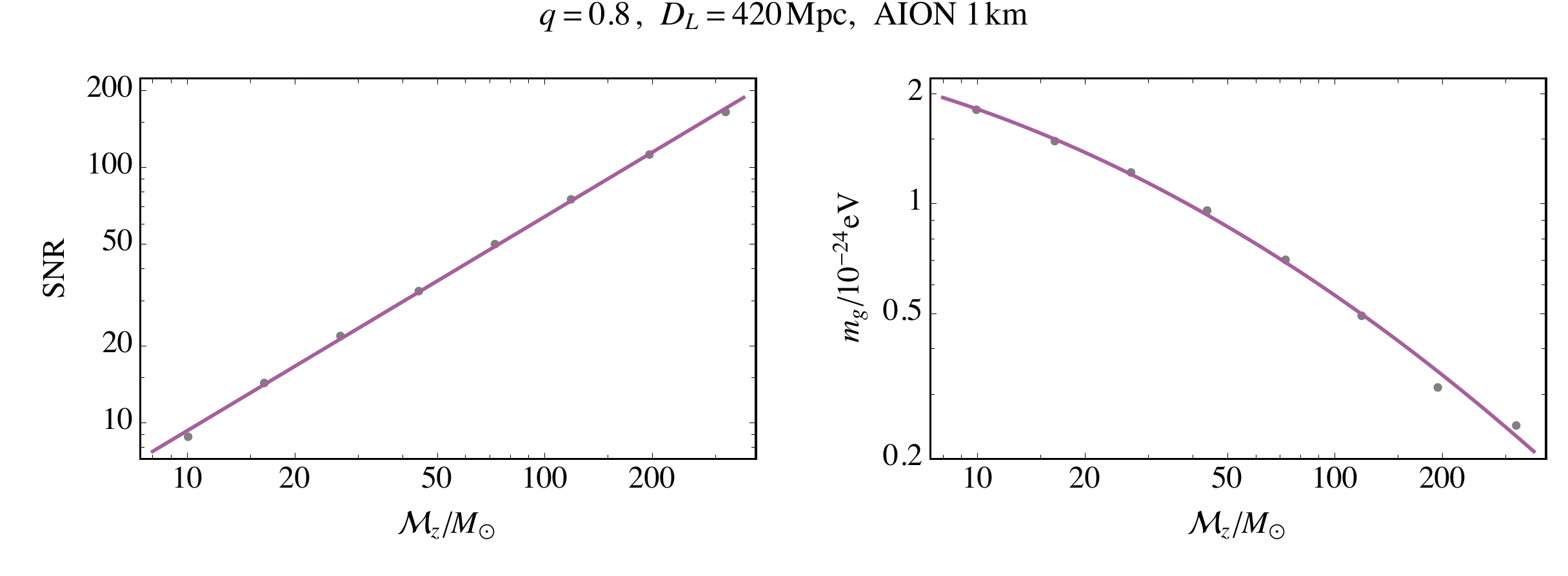}
\includegraphics[width=0.8\textwidth]{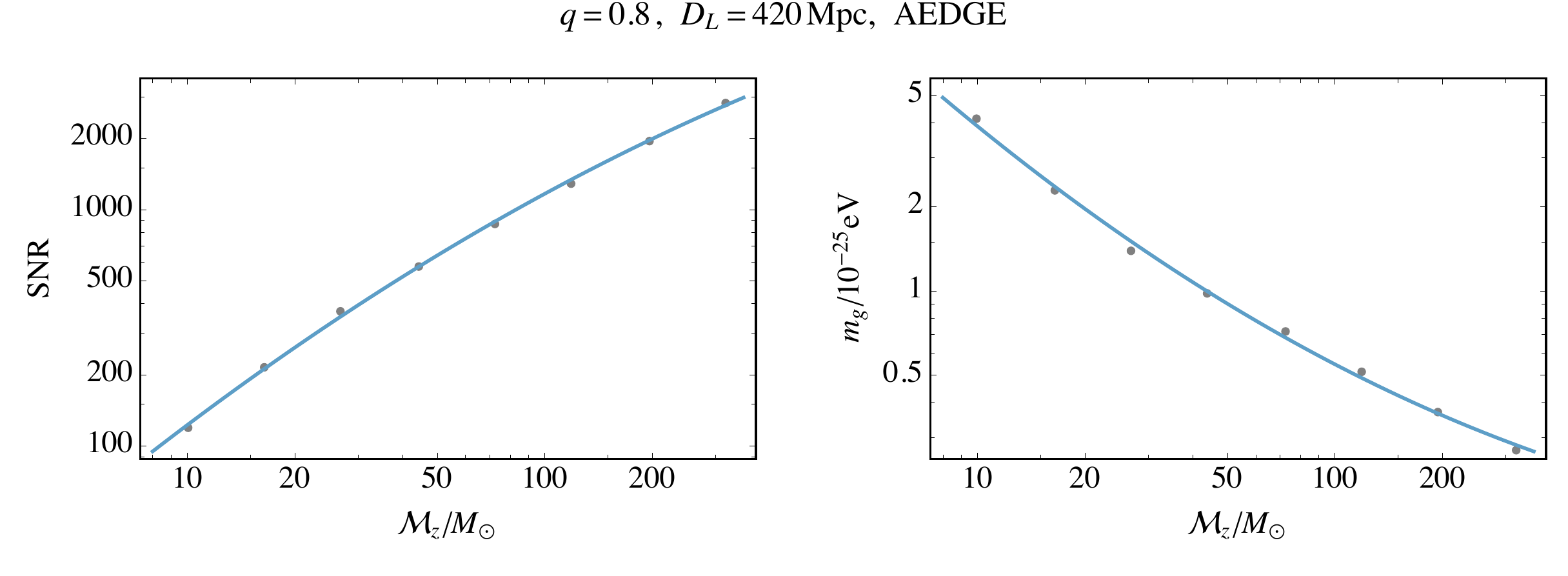}
\vspace{-5mm}
\caption{\it The signal-to-noise ratio (SNR) and the $90\%$ upper bound on the graviton mass for observations of BH-BH binary inspirals as functions of the binary chirp mass for AION\,1km (upper panels) and AEDGE (lower panels).}
\label{fig:BHmasses}
\end{figure*}

We have assumed so far a GW150914-like event at the same luminosity distance $D_L = 420$\,Mpc. However, similar events will certainly occur at larger distances, and also the rate may be such that AION 1\,km and AEDGE could measure GW150914-like mergers at smaller $D_L$. Fig.~\ref{fig:distance} shows in the left panels the SNRs for such an event as a function of $D_L$ between 50 and 1000\,Mpc, and the right panels show the corresponding sensitivities to $m_g$. For each detector, the scaling of the SNR is roughly ${\rm SNR}\propto D_L^{-0.95}$. Then, because the effect of the modified dispersion relation on the waveform increases with the distance to the source, $\delta\Psi \propto D_0$, where $D_0\approx D_L$ at low redshifts, the sensitivity is roughly independent of the luminosity distance to the source. However, we see for the GW150914-like source a slight increase in the sensitivity in the case of AION 1\,km as $D_L$ increases, because the AION 1\,km noise increases as a function of frequency, as seen in Fig.~\ref{fig:strain}, whereas the frequencies of more distant mergers are redshifted to lower frequencies. For AEDGE the opposite behaviour is seen, but the change in the sensitivity to $m_g$ is not substantial in either case.

AION 1\,km and AEDGE are `guaranteed' to measure GW150914-like GW signals. However, both experiments are capable of measuring GWs from mergers of heavier BHs. Therefore, we have also explored the sensitivity to $m_g$ that might be possible with measurements of such mergers. In Fig.~\ref{fig:BHmasses} we show for AION 1\,km (upper panels) and AEDGE (lower panels) how the SNR (left panels) and the sensitivity to $m_g$ (right panels) depend on the chirp mass of a binary with the mass ratio and the luminosity distance of the source fixed to the GW150914 values, $q=0.8$ and $D_L = 420\,$Mpc.~\footnote{For $q=0.8$ and $D_L = 420\,$Mpc the binary merges outside the AION 1\,km frequency range if $\mathcal{M}_z<650\,M_\odot$ and outside the AEDGE frequency range if $\mathcal{M}_z<410\,M_\odot$.} For example, for $\mathcal M_z = 300\,M_\odot$, corresponding to a binary that is roughly 10 times heavier than the one that caused the GW150914 event, the prospective improvement in the 90\% CL upper limit on $m_g$ is to $m_g < 2.5\times 10^{-25}\,$eV for AION 1\,km and $m_g < 3.0\times 10^{-26}\,$eV for AEDGE{, i.e., a factor of $\sim 4$ improvement compared to the GW150914-like signal.} We find that the upper limit on $m_g$ decreases slightly slower than $1/{\rm SNR}$.

AION and AEDGE would also both be able to measure GWs from mergers of much heavier BHs with masses ${\cal O}(10^3$ to $10^5)$ solar masses~\cite{AION,AEDGE},~\footnote{For a review of the observational evidence for intermediate-mass BHs, see~\cite{Mezcua:2017npy}.} which might play roles in the assembly of supermassive black holes in galactic centres~\cite{Woods:2019rqr}. We have made initial studies of the stand-alone AION and AEDGE sensitivities to $m_g$ from the inspiral stages of such mergers, using the example of an intermediate-mass BH binary with $m_1 = 4400\,M_\odot$ and $m_2 = 5600\,M_\odot$. Such a binary merger could be measured by AION 100\,m if it happens sufficiently nearby. For example, the SNR in AION 100\,m from the inspiral phase alone is 7 at $D_L = 420$\,Mpc, and a $90\%$ CL upper {bound $m_g < 1.4\times 10^{-24}$\,eV} could be obtained, even without taking into account measurements of the late stages of the merger. Due to the short observation time, the binary could not be localized, so here we have averaged the signal over $\theta$, $\phi$, $\psi$ and $\cos\tau_i$. Nearby mergers of such heavy BHs would be very rare, but AION 1\,km and AEDGE could see such a binary at much higher redshifts $z \gtrsim 1$ where the rates are more likely to be observable~\cite{Erickcek:2006xc}. If the source redshift is $z=1$, observing the inspiral with AION 1\,km yields ${\rm SNR} = 60$ and $m_g < 8.1\times 10^{-25}$\,eV. At the same distance, AEDGE would yield ${\rm SNR} = 1600$ and $m_g < 8.1\times 10^{-27}$\,eV, and for $z=10$ it would yield ${\rm SNR} = 55$ and $m_g < 4.7\times 10^{-26}$\,eV. Taking the merger-ringdown phase also into account would improve these results, but we see already that if intermediate-mass BH binaries are seen by AION or AEDGE they will improve significantly the bound on the graviton mass.

\subsection{Lorentz violation}

\begin{figure*}
\centering
\includegraphics[width=0.7\textwidth]{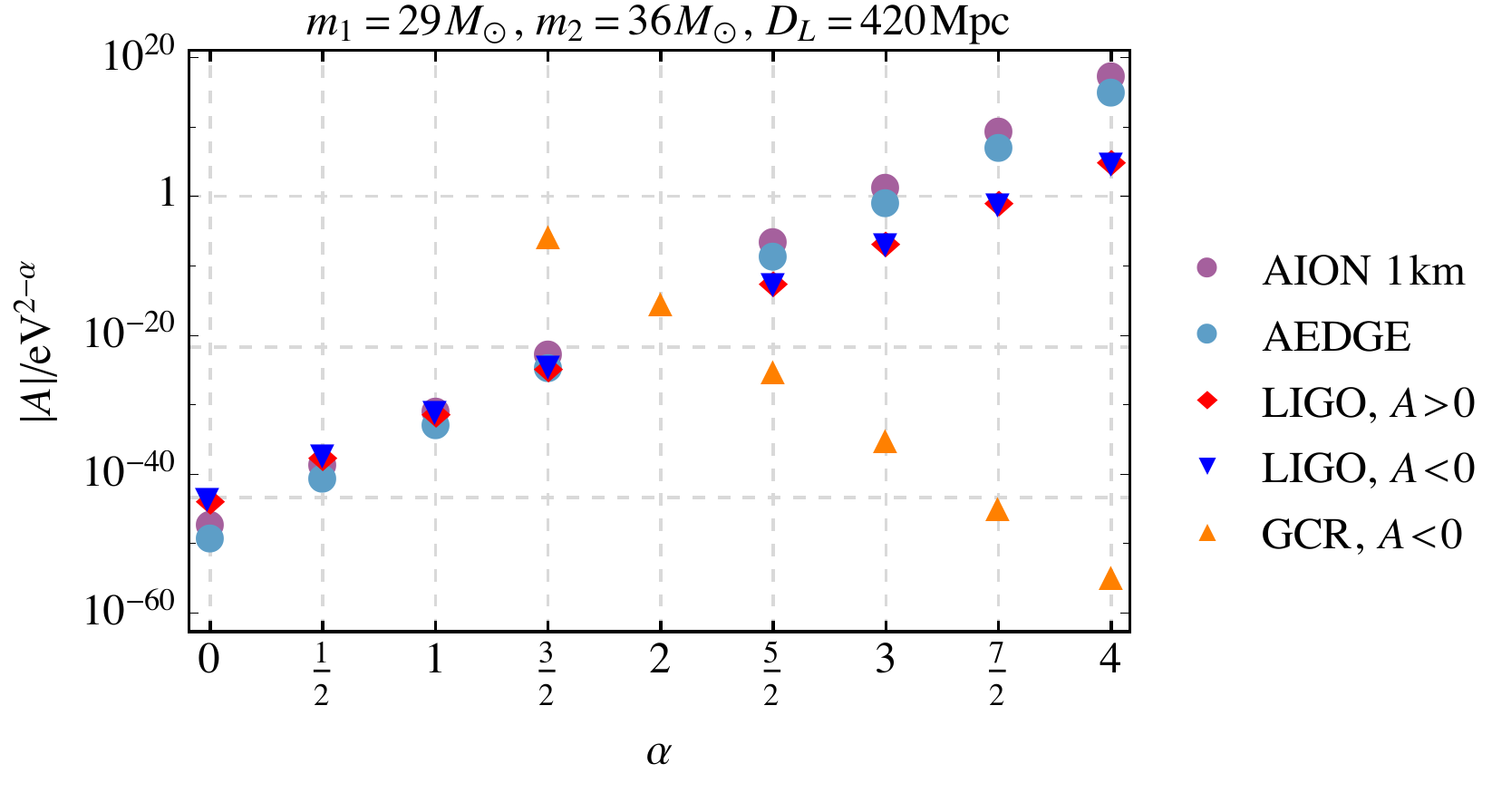}
\caption{\it 90\% CL upper bounds for the magnitude of the modification~\eqref{eq:dispersion} to the relativistic dispersion relation for GWs for different values of $\alpha$. The LIGO constraints are taken from Ref.~\cite{LIGOScientific:2019fpa} and the gravitational Cherenkov radiation (GCR) constraints are taken from Ref.~\cite{Kiyota:2015dla}.}
\label{fig:Aalpha}
\end{figure*}

Modifications of the form of Eq.~\eqref{eq:dispersion} to the GW dispersion relation with $\alpha \ne 0$ violate Lorentz invariance, but are expected in certain modified theories of gravity, e.g., a modification of this form with $\alpha = 3$ was suggested in Ref.~\cite{AmelinoCamelia:1997gz}.~\footnote{For other examples, see, e.g., Ref.~\cite{Mirshekari:2011yq}.} 

In Fig.~\ref{fig:Aalpha} we show the $90\%$ upper bounds on $|A|$ for different values of $\alpha$ as could be obtained by AION and AEDGE observations of the inspiral of a GW150914-like BH binary. The constraint from our analysis does not depend on the sign of $A$. The case $\alpha = 2$ corresponds to a constant difference between the propagation speeds of GWs and of light, which cannot be constrained by GW observations alone. We see in Eq.~\eqref{eq:deltaPsi} that the effect of the modified dispersion relation decreases as a function of frequency if $\alpha < 2$, in which case experiments that are sensitive to lower frequencies probe $A$ more sensitively. Indeed, we see in Fig.~\ref{fig:Aalpha} that the AION 1\,km and AEDGE sensitivities to $A$ are better than that of LIGO for {$\alpha \le 1$}. {Specifically, AION 1\,km is more sensitive than LIGO by a $\mathcal{O}(1000)$ factor for $\alpha = 0$ and by $\mathcal{O}(10)$ for $\alpha = 1/2$, and AEDGE is more sensitive than LIGO by a $\mathcal{O}(10^5)$ factor at $\alpha = 0$, by $\mathcal{O}(1000)$ at $\alpha = 1/2$ and by $\mathcal{O}(10)$ at $\alpha=1$.} On the other hand, the LIGO sensitivity is better for larger $\alpha$, since it is able to measure the later infall stages when the GW frequencies are higher.

The modified dispersion relation~\eqref{eq:dispersion} changes the group velocity of GWs: $v_g = 1+(\alpha-1)AE^{\alpha-2}/2 + \mathcal{O}(AE^{\alpha-2})^2$. In particular, GWs travel slower than the speed of light when $(\alpha - 1) A < 0 $. In this case, if there is no Lorentz violation in their propagation, massive particles can move faster than GWs  and radiate gravitational Cherenkov radiation. The absence of evidence for this effect in cosmic ray observations gives bounds that for $\alpha\geq 2$ and $A<0$ that are stronger than what can be obtained from GW observations~\cite{Kiyota:2015dla}, as also shown in Fig.~\ref{fig:Aalpha}.

\section{Conclusions}

We have analyzed in this paper how the atom interferometer
experiments AION and AEDGE could probe modifications of GR in the propagation of GWs via BH-BH merger measurements in the mid-frequency deciHz range. AION 1\,km and AEDGE measurements of GW150914-like mergers of BHs with chirp masses  ${\cal O}(30)M_\odot$ are `guaranteed'. They would yield constraints on $m_g \sim 10^{-24}$ and $\sim 10^{-25}$~eV, respectively, which are factors $\sim 40$ and $\sim 400$ more sensitive than the current LIGO/Virgo constraint~\cite{LIGOScientific:2019fpa}, thanks to their much longer observations of the inspiral stages (see Fig.~\ref{fig:SNR1}) and their accurate characterizations of the source (see Table~\ref{table:gr}).

The AION 1\,km and AEDGE constraints would be strengthened further by observations of the inspiral phases of more massive BHs: see the right panels of Fig.~\ref{fig:BHmasses} for results for chirp masses $\lsim 300 M_\odot$. It is possible that mergers of intermediate-mass BHs weighing $\sim 10^4 M_\odot$ may also be observable. If their rate is large enough, measurements of their inspirals with AION 100\,m would already be sensitive to $m_g \sim 10^{-25}$~eV. Calculations~\cite{Erickcek:2006xc} give encouragement that the rate of intermediate-mass BH mergers might be observable at redshifts $z \sim 1$, in which case AEDGE measurements of the inspirals would be sensitive to $m_g < 10^{-26}$~eV. Both these sensitivities could be improved by taking into account measurements of the later merger and ringdown stages, which would require numerical studies beyond the scope of this work. Finally, we see in Fig.~\ref{fig:Aalpha} that AION 1\,km and AEDGE will have greater sensitivity than LIGO to models of Lorentz violation $\propto A^\alpha$ with $\alpha \leq 1$.

We conclude that AION and AEDGE both offer significant improvements on the current direct constraints on the graviton mass, and also on some scenarios for Lorentz violation.

\acknowledgments 
This work was supported by the UK STFC Grant ST/P000258/1. J.E. also acknowledges support from the Estonian Research Council grant MOBTT5 and V.V. from the Estonian Research Council grant PRG803.

\bibliography{refs}

\end{document}